# Deep Learning Model of Dock by Dock Process Significantly Accelerate the Process of Docking-based Virtual Screening

Wei Ma[1†], Qin Xie[1,2†], Jianhang Zhang[2], Shiliang Li[1], Youjun Xu[2,3], Xiaobing Deng[3] and Weilin Zhang[2*]

1. Shanghai Key Laboratory of New Drug Design, State Key Laboratory of Bioreactor Engineering, School of Pharmacy, East China University of Science and Technology, Shanghai 200237, China
2. Infinite Intelligence Pharma, Beijing 100083, China
3. College of Chemistry and Molecular Engineering, Peking University, Beijing 100871, China
† Equal contribution

## Abstract

Docking-based virtual screening (VS process) selects ligands with potential pharmacological activities from millions of molecules using computational docking methods, which greatly could reduce the number of compounds for experimental screening, shorten the research period and save the research cost. Howerver, a majority of compounds with low docking scores could waste most of the computational resources. Herein, we report a novel and practical docking-based machine learning method called MLDDM (**M**achince **L**earning **D**ocking-by-**D**ocking **M**odels). It is composed of a regression model and a classification model that simulates a classical docking by docking protocol ususally applied in many virtual screening projects. MLDDM could quickly eliminate compounds with low docking scores and the retained compounds with potential high docking scores would be examined for further real docking program. We demonstrated that MLDDM has a good ability to identify active compounds in the case studies for 10 specific protein targets. Compared to pure docking by docking based VS protocol, the VS process with MLDDM can achieve an over 120 times speed increment on average and the consistency rate with corresponding docking by docking VS protocol is above 0.8. Therefore, it would be promising to be used for examing ultra-large compound libraries in the current big data era.

# 1.Introduction

Drug discovery is a costly, complex, and time-consuming process. According to relevant statistics, the cost for development of drugs approved from 2009 to 2018 raised from $300 million to $2.8 billion, depending on different diseases and methods.[1] Computer-aided drug design (CADD) can significantly accelerate the process of drug discovery. Virtual Screening (VS process) as a typical CADD technique has been widely applied in the early stage of drug discovery for accelerating lead discovery.[2] Docking-based virtual screening aims to computationally place and evaluate small molecule compounds one by one at the binding site of a target protein with three dimensional structure. After continuously optimizing the poses of the compounds, molecular docking explicitly recommend possible potential compounds with high docking scores.

Currently, there are many protein-small molecule docking programs available. A summary of them is listed in Supplimentary (**Supp Table S1**), including their release years, release organizations, descriptions, and licenses[3-19]. Since the forms of scoring functions and the sampling methods they used are quite divergent, for certain system different docking program would yield variant docking results. Applying multiple docking programs sequentially in VS process is quite usual in many publications which is drawn as a well-known funnel like workflow. [20-31] In such docking by docking protocol, the first docking process used in this workflow could be the one with very fast screening speed and the second could be more accurate by somehow slowly. Therefore, the balance between speed and quality of the final result are achieved.

Recently, it has been reported that docking based VS process on ultra-large databases can significantly increase the success rate of finding active hits.[32] Meanwhile the sizes of compound databases also increased significantly. For example, the widely used ZINC database, has exceeded 1.3 billion in 2019 from 700 thousand in 2005.[33-35] Marco Capuccini et al[36] carried out VS process on a large-scale compounds library by running existing docking programs on parallel distributed cloud computational resources. This parallel strategy was based on the Apache Spark engine and achieved an overall efficiency of 87%.

However, only a small proportion of the compounds will have high docking scores, and most of the computational time was wasted on the vast majority of 'low-scoring' compounds. Even the improvement of computational resources make it possible to perform VS process on an ultra-large compound library, it is desireable to identify high-score compounds using reliable machine learning filters in advance as the computational time and resources on docking could be significantly reduced.

P. B. Jayaraj et al[37] developed a classification model based on random forest and applied this model to VS process. It can realize accelerated calculations on GPU, which significantly reduce the running time of the entire process of VS process. Francesco Gentile et al[38] reported the application of a deep learning framework to train classification models on the docking scores of a subset of compounds, then the model was used to predict compounds that have not been docked. Their method realized rapid VS process on a billion compounds library—ZINC-15, which greatly reduced the computational cost of VS on an ultra-large compounds library with an acceptable recall rate.

It would be promising to examine whether a workflow by two tandem models could mimic the above docking by docking protocol and applied for the scenario when the absolute number of compounds to be examined increased in one or two orders. In this paper, we setup such a machine learning model: MLDDM (**M**achince **L**earning **D**ocking-by-**D**ocking **M**odels), by training on two docking programs' results. With comprehensive evaluation of the method, MLDDM was proved to have a good ability to identify known active compounds on 10 specific protein targets. Further evaluation on protein target BRAF showed that the speed of the protocol could gain 120 times increment on average. These results show that MLDDM can quickly filter out compounds with low docking scores with high confidence. It is believed to be a good and practical filter in daily docking based virtual screening protocols with ultra large libraries.

## 2.Methods and Dataset

### 2.1 Model Construction and Usage

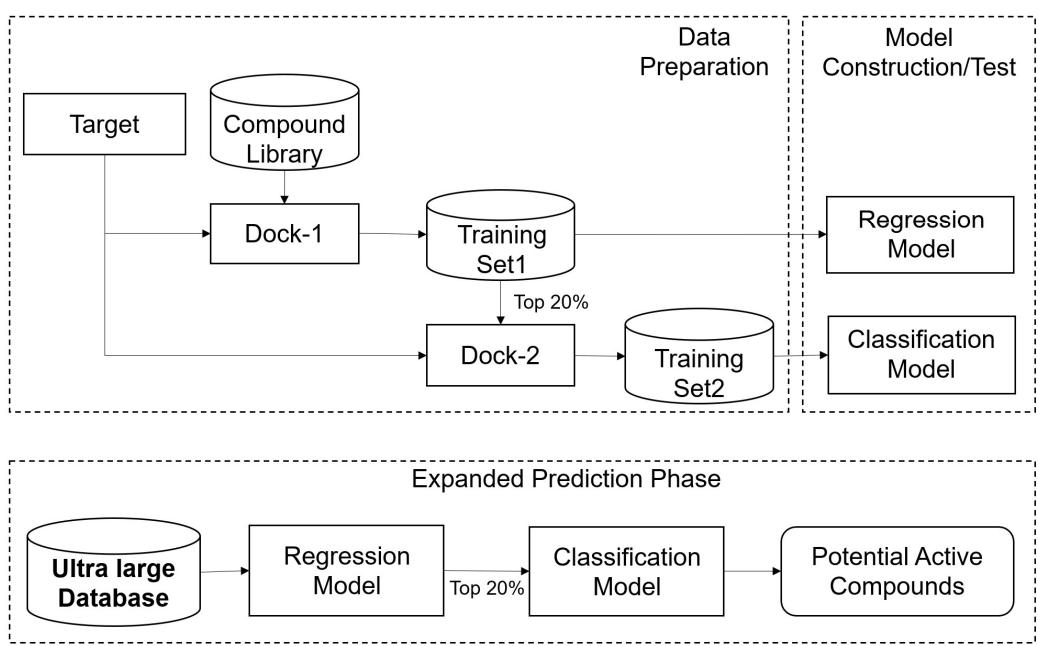

**Figure 1**: The workflow of MLDDM construction and usage.

The framework of the MLDDM model construction and usage is illustrated in **Figure 1**. For certain protein Target, two supervised training processes are implemented. For the first training process, the compounds from the training library (N compounds) were docked into given target. Dock-1 was used as the first docking program, all the docking results are perserved as the first training set (Training Set1) for this target. The compounds with docking score ranking in the top 20% from first docking were selected for second docking process and the docking results from this setup are used as the second training set for this target (Training Set2). We then labeled these training sets acccoding to the docking results. For the second regression task, the rank percentage values of the compounds in Training Set1 are calculated were used as labels. For the classification task, the top 20% ranked compounds within the Training Set2 were selected as the positive samples and the rest compounds were used as the negative samples.The regression model and the classification model were trained seperately. After evaluation of the models, they could be used for ultra-large library filtering. When doing ultra-large library filtering, the regression model are used first which mimics the first step of docking by docking protocol and the high ranking score compounds are selected for the second prediction. The compounds with a positive sign suggested by the classification model could be used for further examination.

## 2.2 Programs and parameter settings for molecular docking

Two open-source and freely available docking programs, namely Vina[39] and rDock[16], were selected for multiple docking VS process in this research. Vina is a molecular docking program based on an experience scoring function developed by the MGL Tools laboratory. Compare with AutoDock 4.0, Vina improves the average accuracy of combined mode prediction, greatly accelerates the search speed by using a simpler scoring function and can still provide reproducibility when dealing with a system with about 20 rotatable bonds. rDock is a fast and versatile Open Source docking program that can be used to dock small molecules against proteins and nucleic acids. Both of the programs can be installed on a computation cluster server and deployed on an almost unlimited number of CPUs, allowing HTVS process campaigns to be carried out in a matter of days.

In our cases, the protein structures were pre-processed using ADTools, and the small molecules were prepared using openbabel.[40] For parameters used in vina, the grid center was defined as the native ligand center in the crystal structure, a cubic grid was used with dimension (*20,20,20*). For rDock, the *rbcavity* program are used to determine a grid parameter from the navtive ligand with default parameters. Vina was selected as the first docking program (Docking-1) and rDock was selected as the second docking program (Docking-2). Both docking results of Docking-1 and Docking-2 were saved as csv files (**Supplementary materials 5**).

## 2.3 Deep learning models

A open souce toolkit, Chemprop (http://github.com/chemprop/chemprop), which is a deep learning based molecular property modeling and prediction toolkit,[41] was used to develop our docking-based machine learning models. The architecture of Chemprop is based on graph convolutional networks, termed as the Directed Message Passing Neural Network (D-MPNN), which treats molecules as an attributed graph with node features (atom) and edge features (bond) for processing. D-MPNN was used to train regression and classification models with basic RDkit molecular descriptors. The default hyperparameters of Chemprop were used for model training. Besides, the setting parameters of

rdkit_2d_normalized and no_features_scaling are added to the training process to improve the stability of the models.

Both the regression and the classification models were trained on 80% of the random split set and then validated and tested on the remaining 20% of the datasets. The hyperparameters of the D-MPNN, such as hidden_size, depth, dropout, and ffn_num_layers were optimized by 10-fold cross validation. The optimal hyperparameters were used to construct the VS-based models. The construction of these models were implemented with Pytorch and RDkit packages.

## 2.4 Datasets and their usages

In this study, several datasets were used for different purposes. Firstly, for the compoud library to be docked into each target, a ChemDiv (ChemDiv Inc) library with 1.25 million purchasable compounds were clustered with a tanimoto similarity threshold 0.4 to obtain a clustered subset of 287,216 compounds (named as ChemDiv subset, and the SMILES file of the compounds are available in **Supplementary materials 2**). Secondly, the description of target proteins used to build the MLDDMs are listed in **Table 1**. The nubmer of their experimentally validated active compounds as well as decoy compounds selected from the DUD-E database[42] are also listed here. They are used to act as validation sets for MLDDM's valibility. All these compounds were docked with two programs to corresponding targets and not used in the training process. (Their smiles strings are list in **Supplementary materials 3**). Thirdly, we tested MLDDM's performance in Expanded prediciton stage by constructing two datasets. For evaluation of those regression models, a relative small dataset are constructed by a subset of 500,000 compounds randomly selected from the ChEMBL database to allievate the actual docking load while keeping the generality of the result. (Their smiles strings are listed in **Supplementary materials 4**) The consistency of the regression model will be evaluated. For evaluation of those classification modes, for each targets, the compounds from the ChEMBL database are labeld as active or inactive depends on whether their $IC_{50}$s are less than 50μM or not.(**Supplementary materials 5**). The whole ChEMBL database (about 19M molecules) were predicted by the regression model, and those passed this filter will be used as the input to the classifcation model. The number of active compounds are evaluated. Finally, ZINC-15 dataset was used to illustrate the speed of MLDDM on a large dataset for target BARF as example.

Table1: Detailed information of the selected targets (from DUD-E) and their active/decoy compounds numbers

| Num | Target | PDB-ID | Type | Actives | Decoys |
|---|---|---|---|---|---|
| 1 | ACE | 3BKL | Protease | 808 | 17144 |
| 2 | ADRB1 | 4BVN | GPCR | 458 | 15958 |
| 3 | BRAF | 3D4Q | Kinase | 251 | 10098 |
| 4 | CDK2 | 1H00 | Kinase | 798 | 28328 |
| 5 | DRD3 | 3PBL | GPCR | 877 | 34188 |
| 6 | DPP4 | 2I78 | Protease | 1079 | 41373 |
| 7 | EGFR | 2RGP | Kinase | 832 | 35442 |
| 8 | JAK2 | 3LPB | Kinase | 153 | 6590 |
| 9 | LCK | 2OF2 | Kinase | 683 | 27856 |
| 10 | VGFR2 | 2P2I | Kinase | 620 | 25280 |

## 3.Results and discussion

### 3.1 Model construction of MLDDM

To reflect the consistency of the model's performance and its generalizability, we summerized the the training statistical metrics of those models for the 10 selected targets in **Figure S1**. The AUC and RMSE on the internal test sets of the D-MPNN models for each target are presented in **Table 2**. For those regression models, the RMSE values ranges from 0.14 to 0.11, illustrating an acceptable error in ranking percentage prediction. For those classification models, the AUC values ranges from 0.89 to 0.96, indicating a good ability to distinguish between positive data (top 20% ranked compounds by Dock-2) and negative data (the rest compounds by Dock-2).

Table 2: Performance on of D-MPNN regression and classification models based on the ChemDIV subset against 10 protein targets.

| Target | Regression RMSE | Classification AUC |
| --- | --- | --- |
| ACE | 0.1237+/-0.0008 | 0.9601+/-0.0030 |
| ADRB1 | 0.1210+/-0.0008 | 0.9478+/-0.0043 |
| BRAF | 0.1323+/-0.0007 | 0.9473+/-0.0047 |
| CDK2 | 0.1177±0.0009 | 0.9448+/-0.0028 |
| DRD3 | 0.1118+/-0.0010 | 0.9478+/-0.0037 |
| DPP4 | 0.1210+/-0.0006 | 0.9257+/-0.0027 |
| EGFR | 0.1235+/-0.0006 | 0.9611+/-0.0035 |
| JAK2 | 0.1340+/-0.0007 | 0.9124+/-0.0039 |
| LCK | 0.1220+/-0.0008 | 0.9604+/-0.0014 |
| VGFR2 | 0.1425+/-0.0005 | 0.9556+/-0.0071 |

## 3.2 Evaluations of regression models

### 3.2.1 Regression model evaluation on DUD-E dataset

Considering there may be some topology bias in DUD-E, we do not intentionally incorporate compounds into training set and use it as one-time validation only.[43] Since the DUD-E dataset also used in succed classifcation model We labeled these as follows: DUD-E actives/decoys were docked by vina and compounds with docking score above the top 20% percentile of the Training-Set-1 were selected. The number of compounds selected were compared with the number of compounds that selected by the regression model. Next, these dock selected compounds were used for the classification model in 3.3.1. The results are listed in **Table 3**. It is shown that among the 10 targets, except for those of ACE, the consistency rates between the docking regression and MLVS classification were 75.0%, with the highest consistency rate of 98% for the CDK2. This suggests that the regression model from MLDDM achieved an acceptable performance as corresponding docking methods.

**Table 3**: Performance of MLDDM regression models and docking results on DUD-E compounds

| Model | Compounds | Number of compounds Selected by Dock-1 top20% | Number of compounds Selected by MLDDM Regression[1] | Consistency rate[3] |
| --- | --- | --- | --- | --- |
| ACE | actives | 55 | 31 | 56.4% |
|  | decoys | 745 | 643 | 86.3% |
| ADRB1 | Actives | 60 | 51 | 85.0% |
|  | decoys | 679 | 575 | 84.7% |
| BRAF | actives | 98 | 88 | 89.8% |
|  | decoys | 949 | 779 | 82.1% |
| CDK2 | actives | 245 | 240 | 98.0% |
| | decoys | 1341 | 1091 | 81.4% |
| DRD3 | actives | 170 | 136 | 80.0% |
| | decoys | 1437 | 1214 | 84.5% |
| DPP4 | actives | 84 | 66 | 78.6% |
| | decoys | 1500 | 1190 | 79.3% |
| EGFR | actives | 225 | 216 | 96.0% |
| | decoys | 2788 | 2308 | 82.8% |
| JAK2 | actives | 34 | 26 | 76.5% |
| | decoys | 395 | 302 | 76.5% |
| LCK | actives | 291 | 253 | 86.9% |
| | decoys | 2096 | 1737 | 82.9% |
| VGFR2 | actives | 158 | 120 | 76.0% |
| | decoys | 1312 | 1086 | 82.8% |

1. Consistency rate equals the Number of compounds Selected by MLDDM Regression over Number of compounds Selected by Dock-1 top20%

### 3.2.2 Regression model evaluation on the ChEMBL subset

We used regression models in MLDDM to predict the ranking percentage of the compoounds in the ChEMBL subset. The compounds with top 20% percentage ranking prediction were further docked to their targets using Vina to test their real distribution.

**Figure 3** shows the superposition of the vina docking score distributions from four representative tagets as examples. The figures for other 6 targets are shown in **Figure S2**. The vina docking score from those selected compounds from ChEMBL subset, the whole ChemDiv dataset and the top 20% scored ChemDiv compounds are illustrated as follows:

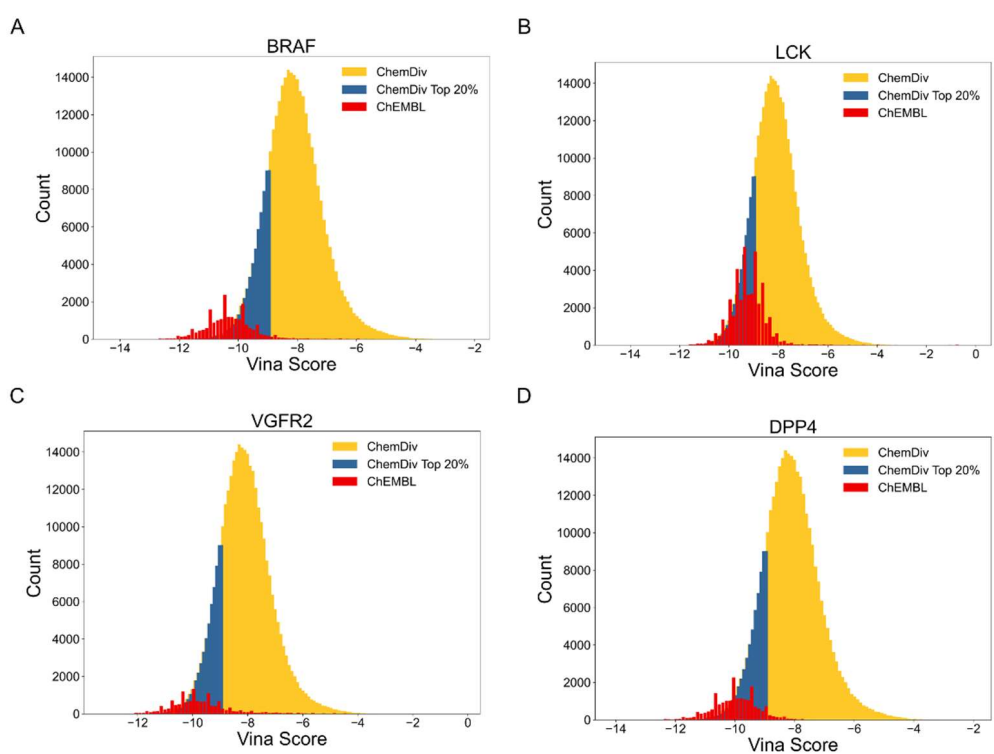

**Figure 3**: Vina docking score distribution for D-MPNN regression model selected ChEMBL compounds and ChemDiv compounds for a) BARF b) LCK c) VGFR2 and d) DPP4.

the yellow bins represent the Vina docking score distribution for ChemDiv, the blue bins represent the top 20% of them, and the red bins represent the vina score distribution from the regression model prediction results of the selected ChEMBL subset compounds.

In **Figure 3**, we can see that most of the selected ChEMBL compounds were located in the top20% docked ChemDiv compounds region. This suggests that the our regression models have a strong ability to enrich compounds with a high docking score, therefore, the models can distinguish compounds with potentially high docking scores for specific targets in a quickly manner without docking.

## 3.3 Evaluation of classification models on the DUD-E sets

In section **3.2.1,** previously docked DUD-E actives/decoys by vina and selected compounds with docking score above the top20% percentile of the Training Set1. These compounds were then docked by rDock. The compounds whose rDock docking score is above the top20% percentile of the Training Set2 was labeled as positive while the rest of them were labeld as negative. These compounds were further predicted by the classification model from MLDDM.

**Table 4** shows the statistical results of the various metrics of the classification model on the DUD-E external data set. The statistical results show that, except for the target DPP4, the accuracies of the other 9 targets on the DUD-E external data set are above 70%. Except for the targets DPP4 and BRAF, the specificity values of the other 8 targets on the DUD-E external data set are above 72%. The false-positive rates of the targets EGFR, DPP4, and BRAF are overall higher than other targets, 24.8%, 46.9%, and 37.9%, respectively, for the rest targets, the false-positive rates are all less than 20%. As for the true positive rate, the maximum and minimum values were 89.5% of BRAF and 27.5% of JAK2. Obviously, for different targets, the performances of the MLDDM are different here. In general, on the external test sets, the models showed a good performance of accuracy and specificity.

**Table 4**: Statistics of various indicators of the model on external data sets

| Num | Target | Accuracy | Precision | Specificity | FPR | TPR | F1-score | Prevalence |
|-----|--------|----------|-----------|-------------|------|------|----------|------------|
| 1 | ACE | 73.9% | 52.0% | 84.0% | 15.9% | 46.3% | 49.0% | 27.1% |
| 2 | ADRB1 | 79.9% | 66.5% | 85.5% | 14.5% | 54.6% | 60.0% | 30.1% |
| 3 | BRAF | 82.8% | 87.9% | 62.1% | 37.9% | 89.5% | 88.7% | 67.6% |
| 4 | CDK2 | 73.3% | 59.3% | 82.2% | 17.8% | 53.8% | 56.4% | 32.2% |
| 5 | DRD3 | 76.0% | 60.2% | 89.7% | 10.3% | 40.4% | 48.3% | 27.8% |
| 6 | DPP4 | 60.4% | 41.8% | 53.1% | 46.9% | 77.2% | 54.3% | 32.5% |
| 7 | EGFR | 72.9% | 49.7% | 75.2% | 24.8% | 66.3% | 56.8% | 27.0% |
| 8 | JAK2 | 74.3% | 46.3% | 89.5% | 10.5% | 27.5% | 34.5% | 24.7% |
| 9 | LCK | 77.5% | 63.4% | 90.3% | 9.7% | 44.1% | 52.0% | 27.6% |
| 10 | VGFR2 | 72.9% | 62.2% | 90.7% | 9.3% | 33.8% | 43.8% | 31.2% |

## 3.4 Overall performance of MLDDM to retrieve all active compounds

To further verify whether the model can identify active compounds for specific targets after the whole MLDDM process, the active compounds of the 10 targets were selected from ChEMBL for verification. The active compounds for 10 targets were filtered and considering a relative loose threshold. If a compound has an activity of $IC_{50}$ less than 50μM, it would be kept as a positive sample.

To explore the models' ability of recalling actives, the whole 19 million molecules from the ChEMBL database were predicted with the regression model from the correponding MLDDM and the compounds which passed the top20% percentile of the training set were selected to make the second prediction. After using the classification model, we count the number of activies recalled.

In **Table 5**, it shows the percentage of recalled compounds in the total positive sample with the different activity cutoff. The statistical results showed that most of the models can identify over 50% of active compounds for specific targets, for example, the proportions of ADRB1, BRAF, EGFR, JAK2, LCK, and VGFR2 were 65.78%, 79.30%, 68.63%, 51.88%, 74.61%, 52.72%, respectively. That is to say, for most targets, the potential active compounds could be recalled by our models.

Table 5: Active compounds prediction statistics of the model

| Target | Threshold of active compounds from ChEMBL ($IC_{50}$) | | | | | | | |
|---|---|---|---|---|---|---|---|---|
| | <50nM | <100nM | <200nM | <500nM | <1μM | <10μM | <20μM | <50μM |
| ACE | 30.00% | 31.56% | 32.73% | 34.46% | 33.42% | 32.35% | 31.83% | 31.10% |
| ADRB1 | 44.62% | 55.81% | 60.68% | 62.89% | 64.44% | 66.81% | 65.34% | 65.78% |
| BRAF | 79.01% | 79.18% | 79.37% | 79.26% | 79.14% | 79.23% | 79.32% | 79.30% |
| CDK2 | 23.29% | 23.08% | 23.34% | 24.22% | 24.35% | 29.41% | 29.41% | 28.80% |
| DPP4 | 34.81% | 34.64% | 34.24% | 33.12% | 32.55% | 32.18% | 32.34% | 32.34% |
| DRD3 | 28.97% | 29.45% | 33.51% | 35.68% | 38.91% | 42.17% | 42.04% | 42.43% |
| EGFR | 74.45% | 74.30% | 73.17% | 72.46% | 69.84% | 68.88% | 68.96% | 68.63% |
| JAK2 | 55.95% | 55.03% | 53.59% | 52.39% | 51.97% | 51.78% | 51.82% | 51.88% |
| LCK | 77.97% | 78.19% | 78.94% | 78.31% | 77.74% | 74.77% | 74.66% | 74.61% |
| VGFR2 | 36.53% | 40.93% | 43.62% | 46.96% | 48.65% | 52.58% | 52.59% | 52.72% |

To test the generalization of our framework, we used the MLDDM trained from current chemDiv training set and calculated $EF_{0.04}$ on ChEMBL. As the current MLDDM would retrieve top4% (0.2 x 0.2 =0.04) of the training set, the enrichment facotr $EF_{0.04}$ could be calculated and evaluated.

The summarized results of the enrichment factors for 10 targets are listed in **Table 6**. From this table, the enrichment factors against 10 targets range from the lowest 5.86 (CDK2) to the highest 17.03 (ADRB1). However, the predictive ability of the model does not show a strong correlation with activity, which may be related to the way our training datasets were acquired.

Table 6: Enrichment factor statistics of 10 targets

| Target | Actives | Total | Recall Comps | True positive | $EF_{4\%}$ |
|---|---|---|---|---|---|
| ACE | 553 | 1941964 | 72638 | 172 | 8.32 |
| ADRB1 | 529 | 1941940 | 74995 | 348 | 17.03 |
| BRAF | 5222 | 1946633 | 94479 | 4141 | 16.34 |
| CDK2 | 1684 | 1943095 | 95447 | 485 | 5.86 |
| DPP4 | 3779 | 1945190 | 87247 | 1222 | 7.21 |
| DRD3 | 337 | 1941748 | 85286 | 143 | 9.66 |
| EGFR | 8232 | 1949643 | 134286 | 5650 | 9.96 |
| LCK | 1713 | 1943124 | 93253 | 1278 | 15.55 |
| JAK2 | 5181 | 1946592 | 108679 | 2688 | 9.29 |
| VGFR2 | 9329 | 1950740 | 71504 | 4918 | 14.38 |

### 3.3 Computation speed comparison

To demonstrate the advantage of our MLDDM models in speed, the target BRAF was selected as an example of the prediction test. The drug-like molecules (containing 981,247,974 compounds) from the ZINC-15 were selected in this case.

Table 7: Comparison of running time

| Num Compounds (k) | Time Cost-Calculate Desciptors (h) | Time Cost-Prediction (h) | Estimated Time of Docking(h)[1] | Recall Compounds | Recall Rate (%) | Times[2] |
|---|---|---|---|---|---|---|
| 5 | 0.065 | 0.005 | 8.20 | — | — | 117 |
| 10 | 0.111 | 0.009 | 16.41 | — | — | 137 |
| 20 | 0.202 | 0.018 | 32.82 | — | — | 149 |
| 50 | 0.509 | 0.041 | 82.04 | 5 | 0.001 | 149 |
| 100 | 1.101 | 0.079 | 164.08 | 11 | 0.001 | 139 |
| 200 | 2.393 | 0.157 | 328.16 | 19 | 0.001 | 129 |
| 500 | 6.009 | 0.391 | 820.40 | 446 | 0.009 | 128 |
| 1,000 | 12.111 | 0.779 | 1640.80 | 1143 | 0.011 | 127 |
| 2,000 | 24.323 | 1.557 | 3281.60 | 2364 | 0.012 | 127 |
| 5,000 | 59.319 | 3.891 | 8204.00 | 8979 | 0.018 | 129 |
| 10,000 | 122.851 | 7.779 | 16408.00 | 33305 | 0.033 | 125 |
| 20,000 | 254.803 | 15.557 | 32816.00 | 111281 | 0.056 | 121 |
| 50,000, | 638.009 | 38.891 | 82040.25 | 686365 | 0.130 | 121 |
| 98,124 | 1249.044 | 77.586 | 161002.35 | 6213291 | 0.633 | 121 |

**Notes:**

Model prediction computing resource: 10 * CPUs (Xeon Gold 5118 ) +1 * GPU(Tesla V100）;

Molecular docking computing resources: 20 * CPUs (Xeon E5-2670 V2);

1. The docking time is linearly predicted according to the docking speed of 300,000 ChemDiv compounds;
2. Speed improvement based on the above resource allocation;

The statistical results in **Table 7** show that based on the above available computing resources, the prediction speed of the machine learning VS model is greatly improved compared to traditional dock by dock process with an average speed increament of >120 times. A large amount of time was used to calculate the descriptors with single-core CPU are included in case every time a unique compound library is used. It means that the MLDDM we built here could greatly increase the speed of dock-by-dock process. This framework could quickly filter out low-scoring compounds against the specific targets, at the same time, the compounds with high docking score were recalled quickly. Therefore, the calculation time and resources of VS on ultra-large compounds library can be greatly reduced.

## 4 Discussion

To explore the reasonable top threshold in the process of obtaining the training dataset, the DUD-E active compounds were selected to make a statistic of the relationship between the enrichment ratio of the active compounds and the top value of the two docking programs for the 10 selected targets respectively. Representative results of 4 targets were listed in **Figure 4**, others are shown in **Figure S3**. The X-axis represents the top value of the compounds sorted by the docking score, and the Y-axis represents the proportion of DUD-E active compounds enriched for the corresponding target. The statistical results showed that there is a logarithmic growth relationship between enrichment ratio and top value, and the enrichment ratio has the maximum profit in the top20% for most of the targets. Besides, the active compounds enrichment ability of rDock is better than Vina overall. In addition to the target DPP4, the enrichment ratios of active compounds in the top20% compounds of the two docking programs are both >20%. It showed that the two selected docking programs have a certain ability to enrich active compounds. Thus, top20% was selected for the preparation of the training dataset.

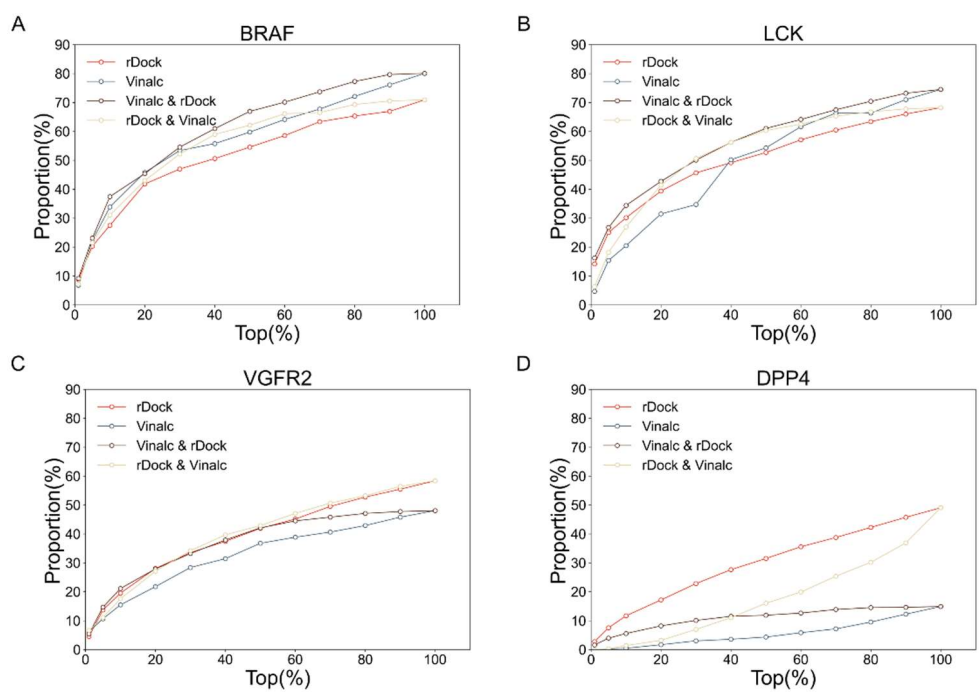

**Figure 4**: Enrichment ratio of DUD-E active compound for two docking programs

To further explore whether the strategy using two docking programs and adjusting their order can obtain a higher enrichment ratio, the enrichment ratio of different combinations of the two docking programs were summarized in **Figure 4**. The red curve in **Figure 4** represents the enrichment ratio of the corresponding target by the rDock program, the blue one represents that of the Vina program, and the purple one represents the re-docking of the Top_20% of the compounds using the rDock after docking with Vina, accordingly, the beige one represents the enrichment ratio in the order of rDock first and then Vina.

The statistical result showed that for most targets, the enrichment ratio of using rDock alone is higher than the other three methods. Besides, choosing to use the Vina program for first docking and then using the rDock program for re-docking of the top20% can make the enrichment ratio further improved. This means that the strategy of combing two docking programs can get more active compounds in the training dataset compared with using Vina alone. Moreover, considering the speed and the consumption of computing resources of the two docking programs, we choose a combination of using Vina for the first step quick screening and then using the rDock program to run a more precise multi-conformations search for re-docking. This strategy ensures that there are enough active compounds in the training dataset we obtained while reducing the time to obtain the training dataset.

Due to the avability of the docking software we could access, we only tested vina and rDock combination which tends to be orthogonal to each other. Conceptionally, this framework should be general to other docking programs' combination and make it very flexible to accommodate different requirements in variant project.

## 5. Conclusion

In the current work, we demonstrated a framework MLDDM which mimics the dock-by-dock screening process powered by deep learning. We implemented it by utilizing the docking results of certain target on a commercially purchasable chemDiv subsets with open access docking program vina and rdock as well as deep learning toolkit Chemprop. Various statistical results showed that the models could have a good performance of VS process. Moreover, compared with the traditional dock by dock process, the MLDDM we constructed here can realize screening on an ultra-large compounds library, and the speed is over 120 times faster. These models can quickly identify molecules with potentially high docking scores and filter out molecules with low docking scores, which greatly reducing the consumption of computing resources on ultra-large compound library virtual screening. In addition, it is found that the reasonable combination of the two docking programs can further increase the enrichment ratio of the screening and make the obtained training dataset more reliable.

With the applications of compound generative models in the field of drug development in recent years, it is believed that the models would be used as a rapid scoring metric for generative models to generate compounds with high docking scores for a specific target.

# Supplementary materials

## Table S1: Summary of common molecular docking programs

| Program | Year Published | Organisation | Description | Webservice | License |
|---|---|---|---|---|---|
| AutoDock[44] | 1990 | The Scripps Research Institute | Automated docking of ligand to macromolecule by Lamarckian Genetic Algorithm and Empirical Free Energy Scoring Function | No | Open source (GNU GPL) |
| AutoDock Vina[39] | 2010 | The Scripps Research Institute | New generation of AutoDock | No | Open source (Apache License) |
| UCSF DOCK[45] | 1988 | University of California-San Francisco | Based on Geometric Matching Algorithm | No | Freeware for academic use |
| FlexX[46] | 2001 | BioSolveIT | Incremental build based docking program | No | Commercial |
| FRED[47] | 2003 | OpenEye Scientific | Systematic,exhaustive,nonstochastic examination of all possible poses within the protein active site combined with scoring Function | No | Freeware for academic use |
| Glide[48] | 2004 | Schrödinger | Exhaustive search based docking program | No | Commercial |
| GOLD[49] | 1995 | Collaboration between the University of Sheffield, GlaxoSmithKlineplc and CCDC | Genetic algorithm based, flexible ligand, partial flexibility for protein | No | Commercial |
| LeDock[50] | 2016 | Lephar | Program for fast and accurate flexible docking of small molecules into a protein | No | Freeware for academic use |
| LigandFit[51] | 2003 | BioVia | CHARMm based docking program | No | Commercial |
| MOE | 2008 | Chemical Computing Group | Docking application within MOE; choice of placement methods (including alpha sphere methods) and scoring functions (including London dG) | No | Commercial |
| PSI-DOCK[14] | 2006 | Peking University | Pose-Sensitive Inclined (PSI)-DOCK | No | Academic |
| rDock[16] | 1998 (commercial)2006 (academic)[14]2012 (open source)[15] | Vernalis R&D(commercial)University of York (academic)University of Barcelona (open source) | HTVS process of small molecules against proteins and nucleic acids, binding mode prediction | No | Open source (GNU LGPL) (formerly commercial, academic) |
| SEED[18] | 1999 | University of Zurich | Automated docking of fragments with evaluation of free energy of binding including electrostatic solvation effects in the continuum dielectric approximation (generalized Born) | No | Open source (GNU GPL) |
| smina[19] | 2012 | University of Pittsburgh | A customized fork of AutoDock Vinawith a better support scoring function and a high-performance energy minimization | No | Open source (Apache License) |
| Surflex-Dock[52] | 2003 | Tripos | Based on an idealized active site ligand (a protomol) | No | Commercial |
| SwissDock[53] | 2011 | Swiss Institute of Bioinformatics | Webservice to predict interaction between a protein and a small molecule ligand | Available | Free to use webservice for academic usage |

# Figure S1

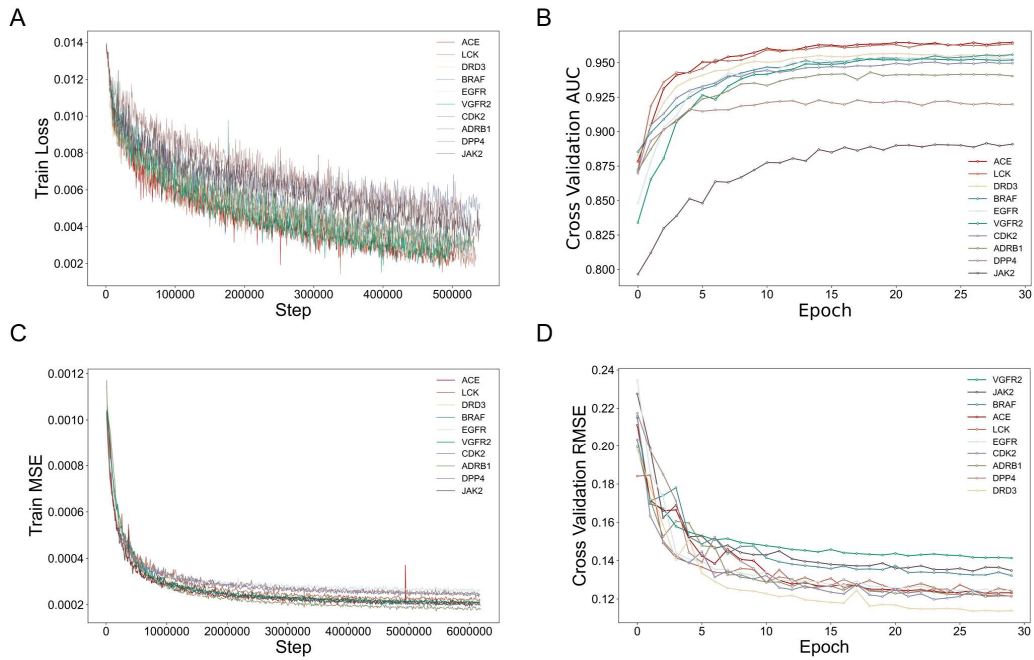

**Figure S1**: The train Loss(classification model), validation AUC(classification model), train MSE(regression model) and validation RMSE(regression model) of 10 selected targets

**Figure S2**

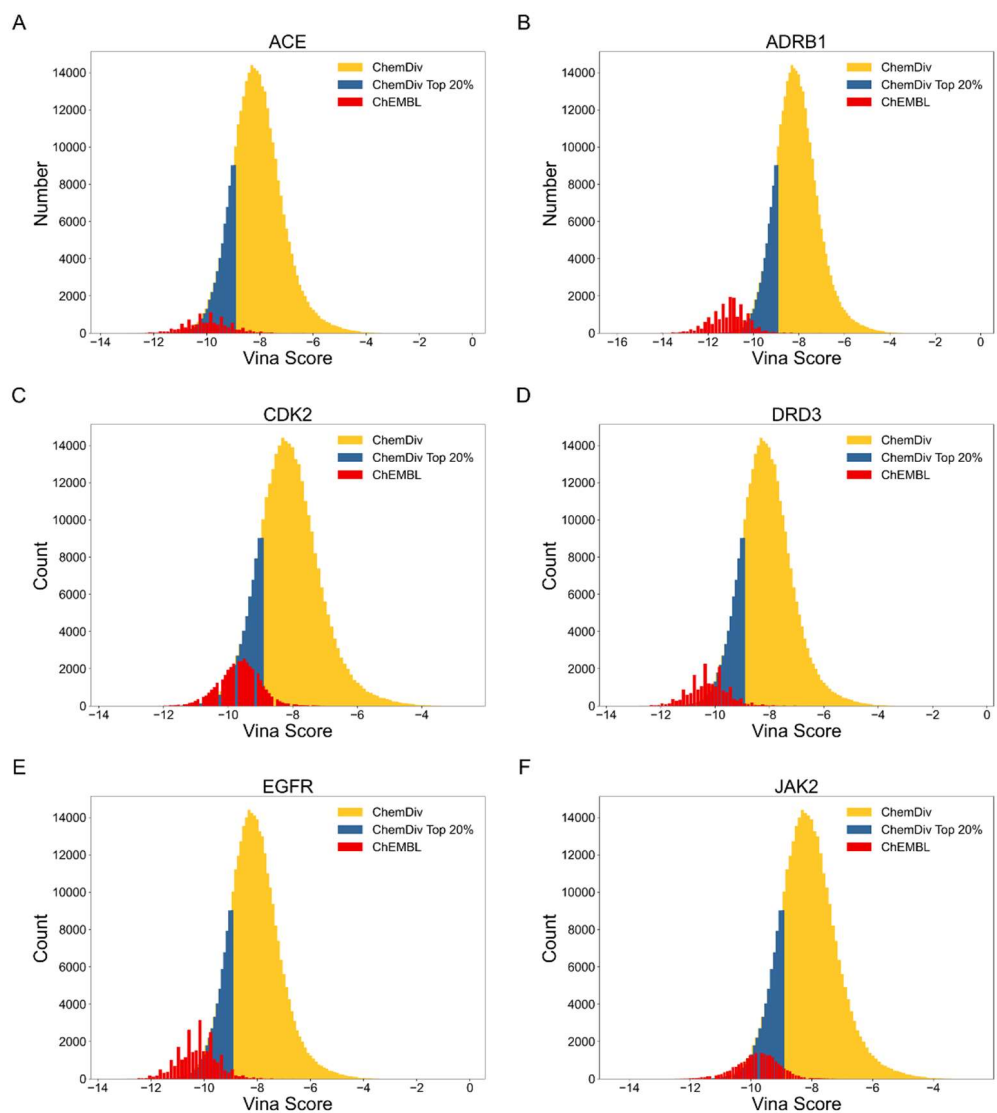

**Figure S2**: Docking score distribution of ChEMBL recalled and ChemDiv compounds.

# Figure S3

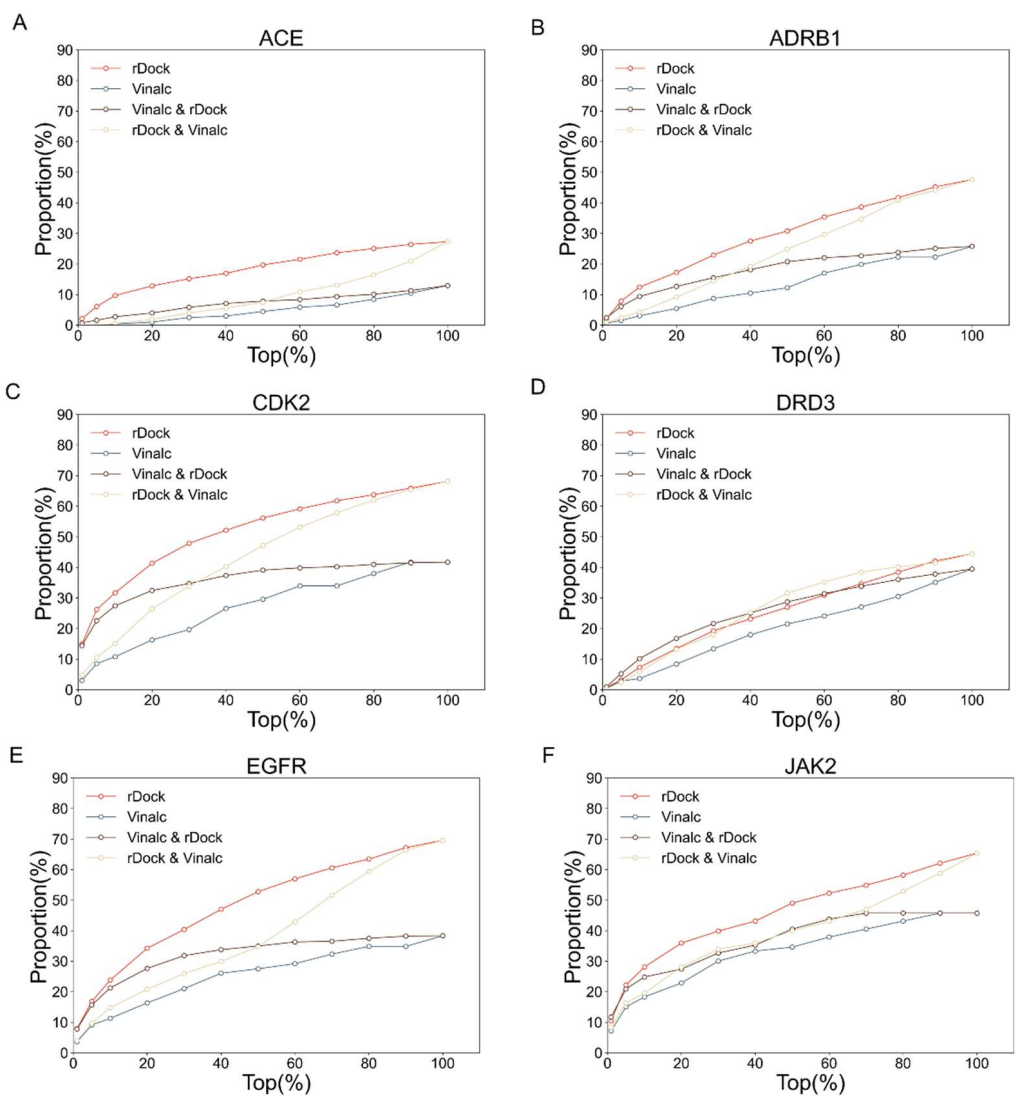

**Figure S3**: Enrichment ratio of DUD-E active compound for two docking programs.